\documentclass[showpacs,amsmath,amssymb]{revtex4}

\usepackage{graphicx}
\usepackage{dcolumn}
\usepackage{bm}
\usepackage{epsfig}

\begin{document}

\def\0#1#2{\frac{#1}{#2}}
\def\bct{\begin{center}} \def\ect{\end{center}}
\def\beq{\begin{equation}} \def\eeq{\end{equation}}
\def\bea{\begin{eqnarray}} \def\eea{\end{eqnarray}}
\def\nnu{\nonumber}
\def\n{\noindent} \def\pl{\partial}
\def\g{\gamma}  \def\O{\Omega} \def\e{\varepsilon} \def\o{\omega}
\def\s{\sigma}  \def\b{\beta} \def\p{\psi} \def\r{\rho}
\def\G{\Gamma} \def\S{\Sigma} \def\l{\lambda}

\title{to understand sQGP through non-topological FL model}
\author{Song~Shu}
\affiliation{Faculty of Physics and Electronic Technology, Hubei
University, Wuhan 430062, China}
\author{Jia-Rong~Li}
\affiliation{Institute of Particle Physics, Hua-Zhong Normal
University, Wuhan 430079, China}
\begin{abstract}
The non-topological FL model is studied at finite temperature and
density. The soliton solutions of the FL model in deconfinement
phase transition are solved and thoroughly discussed for different
boundary conditions. We indicate that the solitons before and
after the deconfinement have different physical meanings: the
soliton before deconfinement represents hadron, while the soliton
after the deconfinement represents the bound state of quarks which
leads to a sQGP phase. The corresponding phase diagram is given.
\end{abstract}  \maketitle

\section{Introduction}
An important discovery at Relativistic Heavy Ion Collider (RHIC)
in recent years is the realization of strongly interacting Quark
Gluon Plasma (sQGP)~\cite{ref1,ref2,ref2a}. For the collective
effects from RHIC experiments, known as radial and elliptic flow,
the QGP could be well described by ideal hydrodynamics. It implies
that the QGP at RHIC is the most perfect
fluid~\cite{ref3,ref4,ref5}. For this reason, Shuryak has pointed
out there should be lots of bound states~\cite{ref1} especially
for light and heavy qq bound states at $T_c<T<4T_c$, where $T_c$
is the transition temperature. For the physical quark mass the QCD
transition is a non-singular crossover at finite temperature. In
recent lattice results, at high temperatures and small densities,
it is a crossover from hadronic phase to QGP
phase~\cite{ref6,ref7}. From recent theoretical studies, we have
realized the rich phases of QCD theory, such as color
superconducting phase~\cite{ref8}, pion condensation~\cite{ref9},
color glass condensate and quarkyonic phase~\cite{ref10}. Now the
RHIC results tell us that after deconfinement, the quarks are not
free immediately, as we usually thought for a weakly coupled QGP
(wQGP), but still in strong coupled state which leads to a new
state of nuclear matter as the sQGP. How to understand the
formation mechanism of sQGP is an important problem~\cite{ref11}.
As we know, there are confine potential and color coulomb
potential between quarks in vacuum, know as $V\sim \alpha/r+kr$.
With temperature $T$ increased to some critical temperature, the
confine potential disappears, while the coulomb potential remains,
which could still be very strong~\cite{ref12}. Thus in
literatures, the effective coulomb potential has been often used
to describe the strong interaction of sQGP~\cite{ref12a}. But from
the theoretical point of view, it is lack of a more fundamental
theoretical model to understand these bound states of quarks. In
this paper we wish to use non-topological FL model to study this
problem.

FL model has been widely discussed in past
decays~\cite{ref13,ref14,ref15}. It has been very successful in
describing phenomenologically the static properties of hadrons and
their behaviors at low energy. The model consists of quark fields
interacting with a phenomenological scalar field $\s$. The $\s$
field is introduced to describe the complicated nonperturbative
features of QCD vacuum. It naturally gives a color confinement
mechanism in QCD theory. The model has been also extended to
finite temperatures and densities to study deconfinement phase
transition~\cite{ref16,ref17,ref18,ref19}. However it seems that
the deep meaning of the soliton solutions in deconfinement phase
transition has not been revealed in the past studies. The main
purpose of this paper is to study in detail the properties of the
solitons in FL model before and after deconfinement, and provide a
natural explanation and description of sQGP through this effective
theoretical model.

The organization of this paper is as follows: in section 2 we give
a brief introduction of FL model. The field equations and the
effective potential are derived. In section 3, the soliton
equation of FL model is solved for different boundary conditions
and the physical meanings of these solitons in deconfinement are
thoroughly discussed. In section 4, a phase diagram of
deconfinement phase transition is given. The last section is the
summary.

\section{FL model and the equations}
We start from the Lagrangian of the FL model, \bea {\cal
L}=\bar\psi(i\gamma_\mu\pl^\mu-g\s)\psi+\012(\pl_\mu\s)(\pl^\mu\s)-U(\s),
\eea where\bea U(\s)=\01{2!}a\s^2+\01{3!}b\s^3+\01{4!}c\s^4+B.
\eea $\p$ represents the quark field, and $\s$ denotes the
phenomenological scalar field. $a, b, c, g$ and $B$ are the
constants which are generally fitted in with producing the
properties of hadrons appropriately.

For the thermal equilibrium system, the $\s$ field will be
replaced by $\bar\s+\s'$, where $\bar\s$ is the Gibbs thermal
average of the $\s$ field, and $\s'$ is the fluctuation. At mean
field approximation, one can obtain the field equations as
follows, \bea &&(i\gamma_\mu\pl^\mu-g\bar\s)\psi=0,
\\ &&\pl_\mu\pl^\mu\bar\s=-\0{\pl
V_{eff}}{\pl\bar\s}, \label{mean} \eea where $V_{eff}$ is one loop
effective potential at finite temperature and density. Notice that
at mean field approximation the quantum corrections have been
neglected. The $V_{eff}$ could be derived through finite
temperature field theory~\cite{ref18,ref19}, \bea
V_{eff}=U(\bar\s)+\01{\b}\int \0{d^3\bf p}{(2\pi)^3}\ln (1-e^{-\b
E_{\s}})-\0{\g}{\b}\int \0{d^3\bf p}{(2\pi)^3}\left[\ln
(1+e^{-\b(E_q-\mu)}) + \ln (1+e^{-\b(E_q+\mu)})\right],
\label{eff} \eea in which $\b$ is the inverse temperature, $\mu$
is the chemical potential, and $\g$ is a degenerate factor,
$\g=2(spin)\times 2(flavor)\times 3(color)$. Besides
$E_\s=\sqrt{\vec p^2+m_\s^2}$, $E_q=\sqrt{\vec p^2+m_q^2}$, where
$m_q=g\bar\s$ and $m_\s^2=a+b\bar\s+\012c\bar\s^2$ are the
effective masses of the quark and $\s$ field respectively.

From equation (\ref{mean}), one could see that the properties of
the soliton field $\bar\s$ depend completely on the effective
potential $V_{eff}$. As known, by the effective potential at
finite temperature and density, one can study the deconfinement
phase transition in FL model. Thus the properties of solitons in
deconfinement, especially the relations between solitons and
deconfinement, could be well studied through solving the equation
(\ref{mean}) at finite temperature and density.

\begin{figure}[tbh]
\begin{center}
\includegraphics[width=210pt,height=150pt]{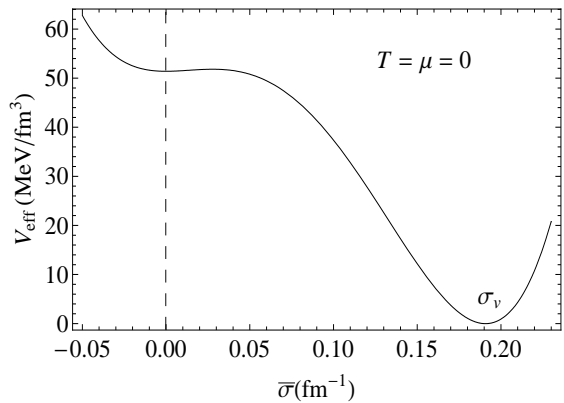}
\end{center}
\caption{The effective potential at zero temperature and density
.}\label{f1}
\end{figure}

In our calculation, the parameters are chosen to be
$a=17.7fm^{-2}, b=-1457.4fm^{-1}, c=20000, g=12.16$. The effective
mass of $\s$ field is fixed at $m_\s=550MeV$~\cite{ref18}. At zero
temperature and density, the $V_{eff}$ is just the $U(\bar\s)$ and
could be plotted in Fig.\ref{f1}. There are two minima: one
corresponds to the perturbative vacuum at $\bar\s=0$, another
corresponds to the physical vacuum at $\bar\s=\s_v$.

\section{The soliton solutions and sQGP in deconfinement}
For the static and spherically symmetric soliton field, the
equation (\ref{mean}) becomes \bea
\0{d^2\bar\s}{dr^2}+\02r\0{d\bar\s}{dr}=\0{\pl
V_{eff}}{\pl\bar\s}. \label{soliton} \eea We will solve this
equation numerically. A standard numerical package COLSYS will be
used for the numerical calculation~\cite{ref20}.

For solving the soliton equation (\ref{soliton}), one should first
determine the boundary condition. There are two different cases of
the boundary conditions for solving the
equation~\cite{ref15,ref21}. In the following the two cases will
be studied separately.

\subsection{Boundary condition I}
In this case, we may consider the mechanical analog in which there
is a point particle at ``positon" $\s$ and a time ``r", moving in
a ``potential" $-V_{eff}$. The discussions about this mechanical
analog could be found in Ref.~\cite{ref21}. Equation
(\ref{soliton}) is then the movement equation of the particle.
Because the energy conservation will be applied in this case, the
second term at l.h.s of equation (\ref{soliton}) will be omitted.
Thus the equation becomes \bea \0{d^2\bar\s}{dr^2}=\0{\pl
V_{eff}}{\pl\bar\s}. \label{soliton1} \eea In order to determine
the boundary condition, suppose at ``time" $r=-\infty$, the
``position" of the particle is at $\bar\s=0$. From Fig.\ref{f1},
if the particle is pushed very gently to start moving, the
particle will move to another point just at $\bar\s=\s_a$ where
$V_{eff}(0)=V_{eff}(\s_a)$, noticing that the particle is moving
in the ``potential" $-V_{eff}$. Because of energy conservation,
the particle will then move back and return to the point
$\bar\s=0$. Thus one could obtain the following boundary
condition, \bea \bar\s(r=-\infty)=0, \ \ \ \ \bar\s(r=+\infty)=0.
\eea  From Fig.\ref{f1}, the physical vacuum $\bar\s=\s_v$ is
stable, and $B=V_{eff}(0)-V_{eff}(\s_v)$ is the bag constant. The
quarks are confined in a soliton bag. At zero temperature and
density, the soliton solution represents a hadron.

\begin{figure}[tbh]
\begin{center}
\includegraphics[width=210pt,height=150pt]{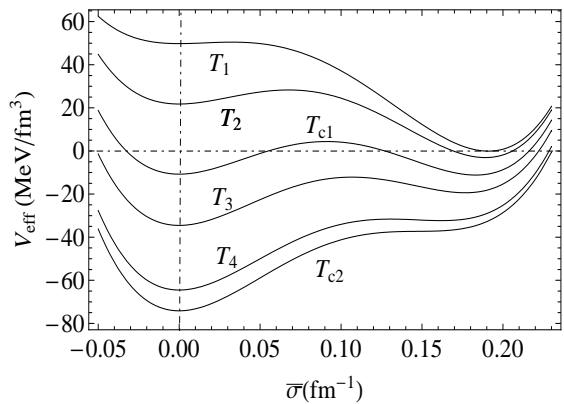}
\end{center}
\caption{The effective potential for different temperatures:
$T_1=50MeV$, $T_2=100MeV$, $T_{c1}=121MeV$, $T_3=130MeV$,
$T_4=140MeV$ and $T_{c2}=143MeV$.}\label{f2}
\end{figure}

At finite temperature, the $V_{eff}$ could be plotted in
Fig.\ref{f2}. There are two critical temperatures $T_{c1}$ and
$T_{c2}$. At $T_{c1}$, the two minima degenerate and the bag
constant $B=0$. At $T_{c2}$, the minimum $\bar\s=\s_v$ just
vanishes. For different configurations of $V_{eff}$, the equation
(\ref{soliton1}) could be numerical solved by the numerical
package COLSYS. However the $V_{eff}$ is highly non-linear in
$\bar\s$ as shown in equation (\ref{eff}). It could not be
directly used in COLSYS evaluation. At finite temperature, as the
$V_{eff}$ could be plotted as shown in Fig.\ref{f2}, we can fit
the curve in terms of $\s^2$, $\s^3$ and $\s^4$ with the effective
parameters $a(T)$, $b(T)$ and $c(T)$. That means \bea
V_{eff}=\01{2!}a(T)\s^2+\01{3!}b(T)\s^3+\01{4!}c(T)\s^4+B(T). \eea
In this form, the $V_{eff}$ could be applied in COLSYS for the
numerical calculation. In the following we will solve equation
(\ref{soliton1}) and discuss the soliton solutions for different
configurations of $V_{eff}$.

\begin{figure}[tbh]
\includegraphics[width=210pt,height=150pt]{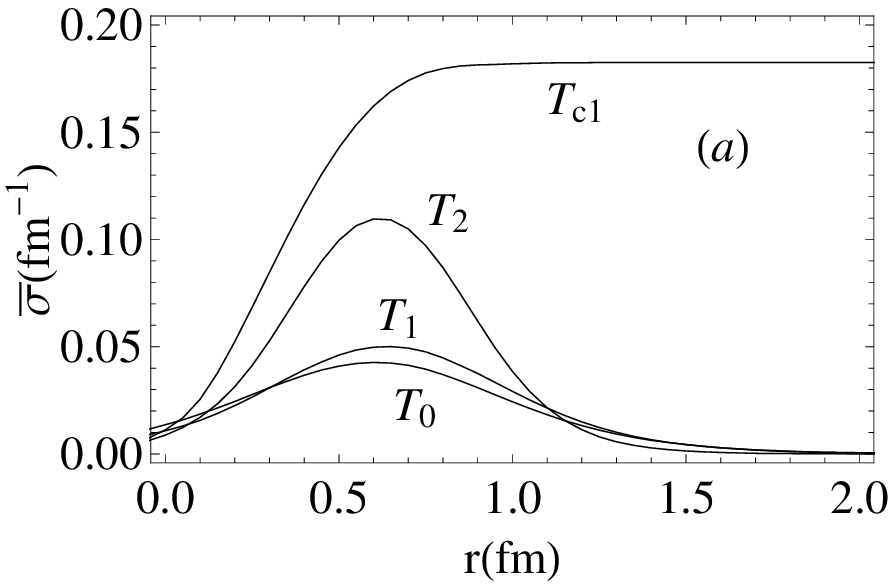}
\hspace{1cm}
\includegraphics[width=210pt,height=150pt]{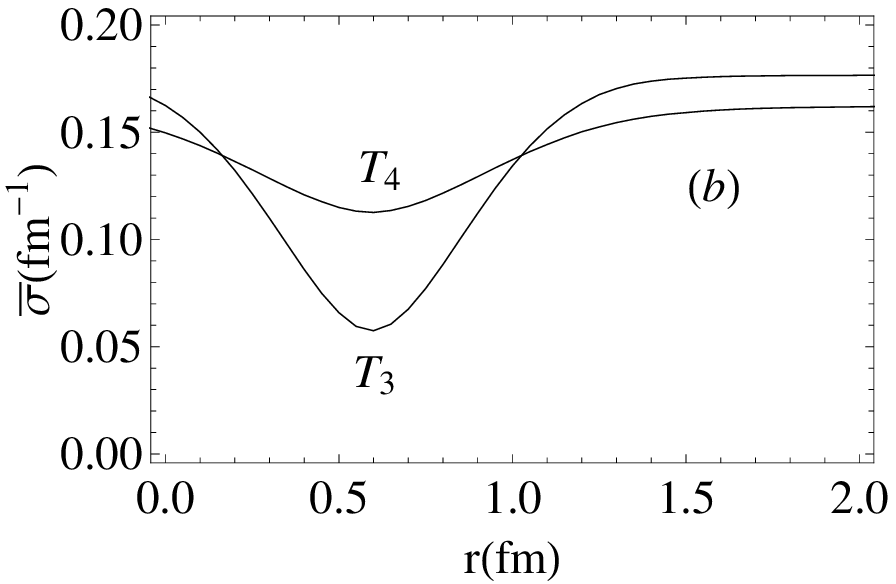}
\caption{The soliton solutions for boundary condition I at
different temperatures. (a) $T_0=0MeV$, $T_1=50MeV$, $T_2=100MeV$
and $T_{c1}=121MeV$. (b) $T_3=130MeV$ and $T_4=140MeV$.}\label{f3}
\end{figure}

At $T<T_{c1}$, the physical vacuum at $\bar\s=\s_v$ is stable. The
bag constant $B\neq 0$ and its value decreases with temperature
increasing. The corresponding quarks are confined. The equation
(\ref{soliton1}) could be numerically solved and the soliton
solutions are plotted in Fig.\ref{f3}a. One could see that the
solitons at temperature $T<T_{c1}$ do not change so much when
compared to that at zero temperature except the peaks of the
solitons become higher when temperature increases. The quarks keep
confined in a hadronic state until the critical temperature
$T=T_{c1}$ the two vacuums degenerate. At this time, still
required by the energy conservation the boundary condition is \bea
\bar\s(r=-\infty)=0, \ \ \ \ \bar\s(r=+\infty)=\s_v. \eea The
soliton becomes topological as shown in Fig.\ref{f3}a. As the bag
constant $B=0$ at this time, the hadrons are destructed and the
quarks are deconfined.

If we keep increasing temperature to $T_{c1}<T<T_{c2}$, from
Fig.\ref{f2}, the perturbative vacuum $\bar\s=0$ is stable. The
equation (\ref{soliton1}) will be solved under the boundary
condition \bea \bar\s(r=-\infty)=\s_v, \ \ \ \
\bar\s(r=+\infty)=\s_v. \eea The soliton solutions could be
plotted in Fig.\ref{f3}b. The soliton solutions still exist but
become very different from those at $T<T_{c1}$. The peaks of the
solitons become downward. When temperature increasing, they become
flat. These solitons do not represent hadrons any more but the
bound states of quarks. Though the quarks are deconfined, they are
still in strong coupled states. That is to say the system is in a
sQGP phase. Until $T\geq T_{c2}$, the physical vacuum
$\bar\s=\s_v$ vanishes and only the perturbative vacuum $\bar\s=0$
exists, has the soliton solution disappeared. At this time the
quarks become free. The system goes into a quasi-free gas phase of
QGP.

\begin{figure}[tbh]
\begin{center}
\includegraphics[width=210pt,height=150pt]{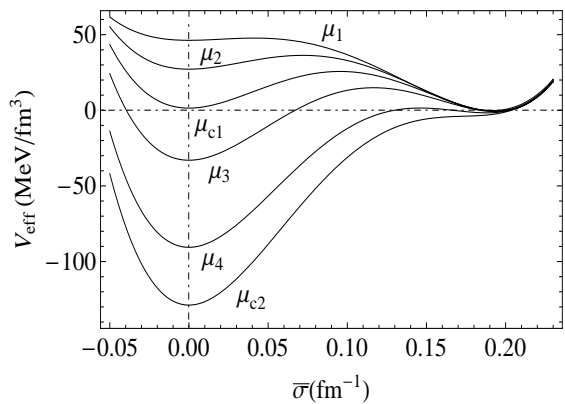}
\end{center}
\caption{The effective potential for different chemical
potentials: $\mu_1=100MeV$, $\mu_2=200MeV$, $\mu_{c1}=255MeV$,
$\mu_3=300MeV$, $\mu_4=350MeV$ and $\mu_{c2}=375MeV$.}\label{f4}
\end{figure}

\begin{figure}[tbh]

\includegraphics[width=210pt,height=150pt]{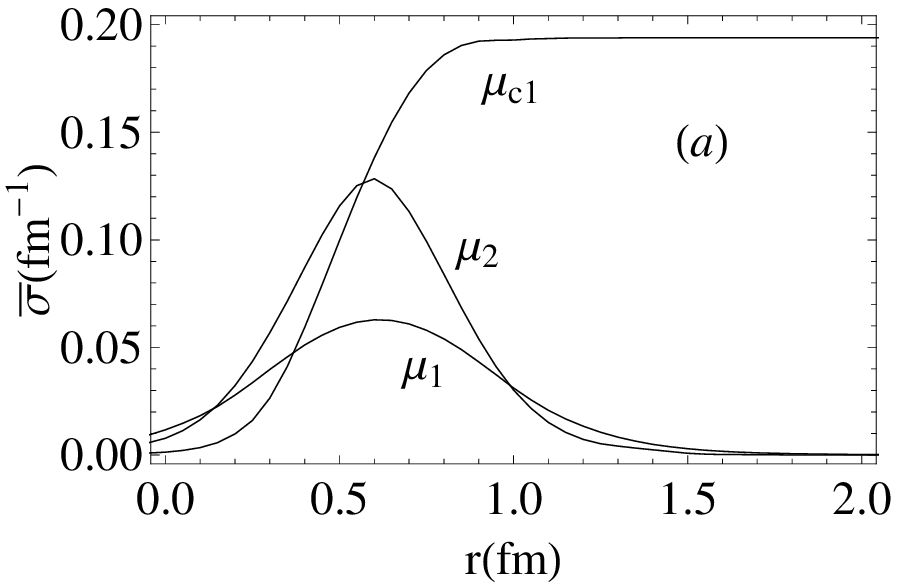}
\hspace{1cm}
\includegraphics[width=210pt,height=150pt]{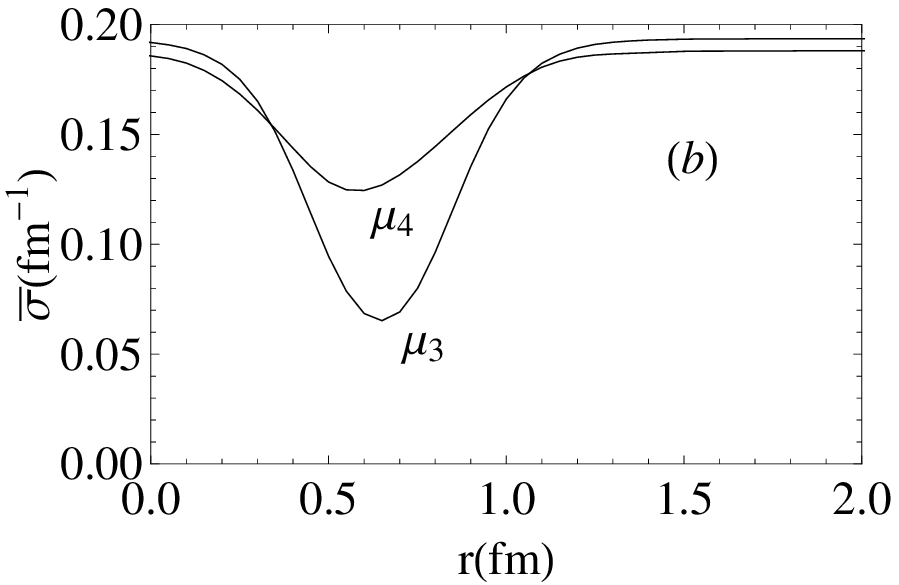}
\caption{The soliton solutions for boundary condition I at
different chemical potentials. (a) $\mu_1=100MeV$, $\mu_2=200MeV$
and $\mu_{c1}=255MeV$. (b) $\mu_3=300MeV$ and
$\mu_4=350MeV$.}\label{f5}
\end{figure}

At finite temperature and density, we could also make the similar
discussions. In Fig.\ref{f4}, we show the $V_{eff}$ at $T=50MeV$
and different chemical potentials. There are also two critical
chemicals $\mu_{c1}$ and $\mu_{c2}$. At $\mu_{c1}$ the two vacuums
degenerate; at $\mu_{c2}$ the vacuum of $\s\neq 0$ just vanishes.
The corresponding soliton solutions are plotted in Fig.\ref{f5}a
and Fig.\ref{f5}b. It is clear that at $\mu\leq\mu_{c1}$, the
system is in a hadronic phase; at $\mu_{c1}<\mu<\mu_{c2}$, the
system is in a sQGP phase; at $\mu\geq\mu_{c2}$, the system
becomes quasi-free gas of QGP.

\subsection{Boundary condition II}
\begin{figure}[tbh]

\includegraphics[width=210pt,height=150pt]{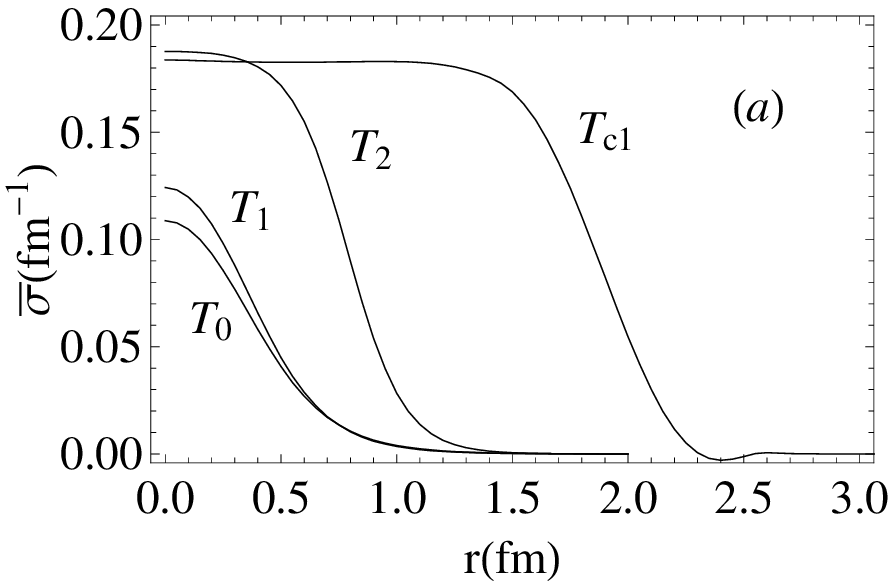}
\hspace{1cm}
\includegraphics[width=210pt,height=150pt]{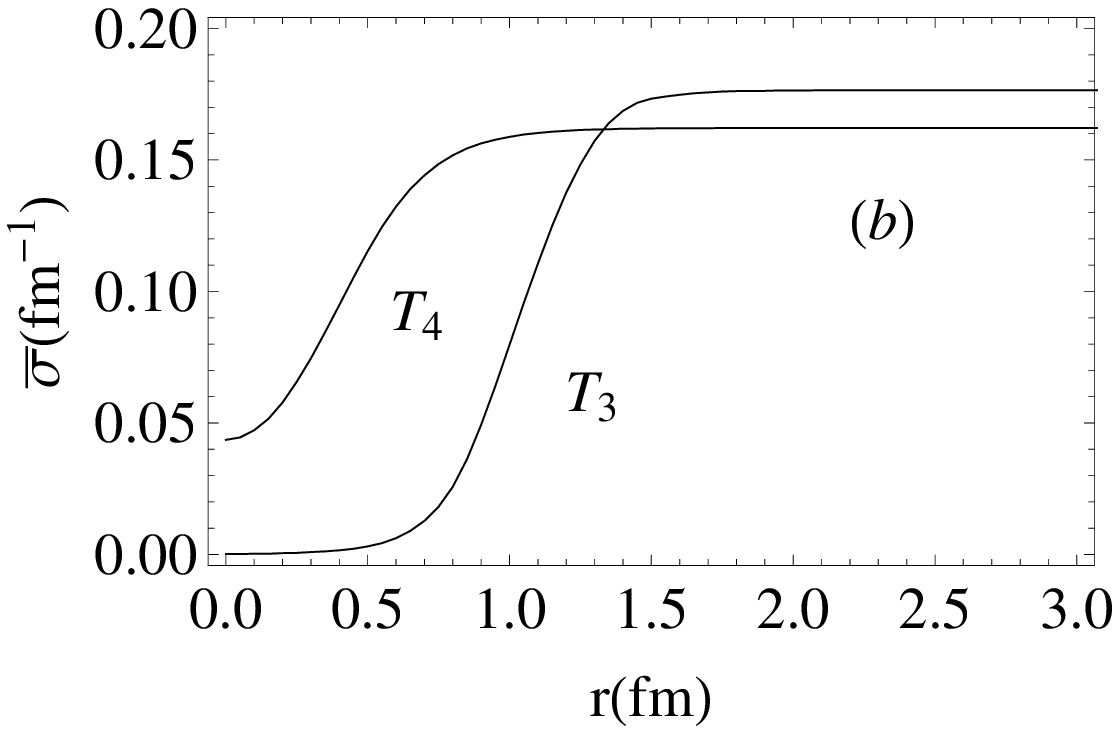}
\caption{The soliton solutions for boundary condition II at
different temperatures. (a)$T_0=0MeV$, $T_1=50MeV$, $T_2=100MeV$
and $T_{c1}=121MeV$. (b) $T_3=130MeV$ and $T_4=140MeV$.}\label{f6}
\end{figure}

In this case, at $T\le T_{c1}$, the boundary condition is taken as
\bea \left.\0{d\bar\s}{dr}\right|_{r=0}=0, \ \ \ \
\left.\bar\s\right|_{r\rightarrow \infty}=0. \eea The equation
(\ref{soliton}) could be numerically solved at finite temperature
and density for different configurations of $V_{eff}$.

From Fig.\ref{f2}, at $T<T_{c1}$, the soliton solutions are
plotted in Fig.\ref{f6}a. The physical vacuum $\bar\s=\s_v$ is
stable and The bag constant $B\neq 0$. The solitons still
represent that the quarks are confined and the system is in a
hadronic phase. Until $T=T_{c1}$ the system are deconfined. The
corresponding soliton is also plotted in Fig.\ref{f6}a.

At $T_{c1}<T<T_{c2}$, the boundary condition is \bea
\left.\0{d\bar\s}{dr}\right|_{r=0}=0, \ \ \ \
\left.\bar\s\right|_{r\rightarrow \infty}=\s_v. \eea The soliton
solutions are solved and plotted in Fig.\ref{f6}b. These solitons
are quite different from those at $T<T_{c1}$. As the system is
already deconfined, these solitons represent the quarks are in a
bound state. The system is in a sQGP phase. At $T>T_{c2}$, there
are no soliton solution any more, the system goes to a quasi-free
gas of QGP.

\begin{figure}[tbh]

\includegraphics[width=210pt,height=150pt]{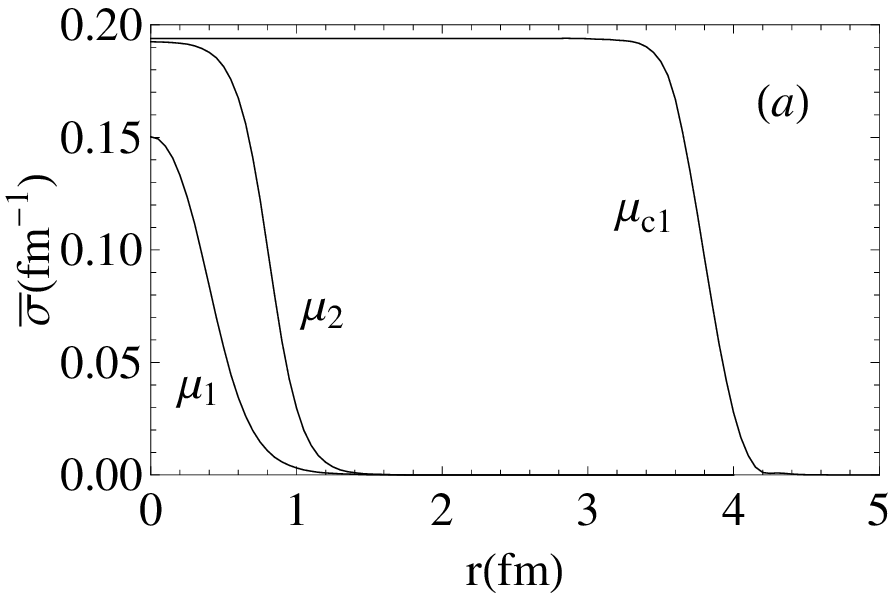}
\hspace{1cm}
\includegraphics[width=210pt,height=150pt]{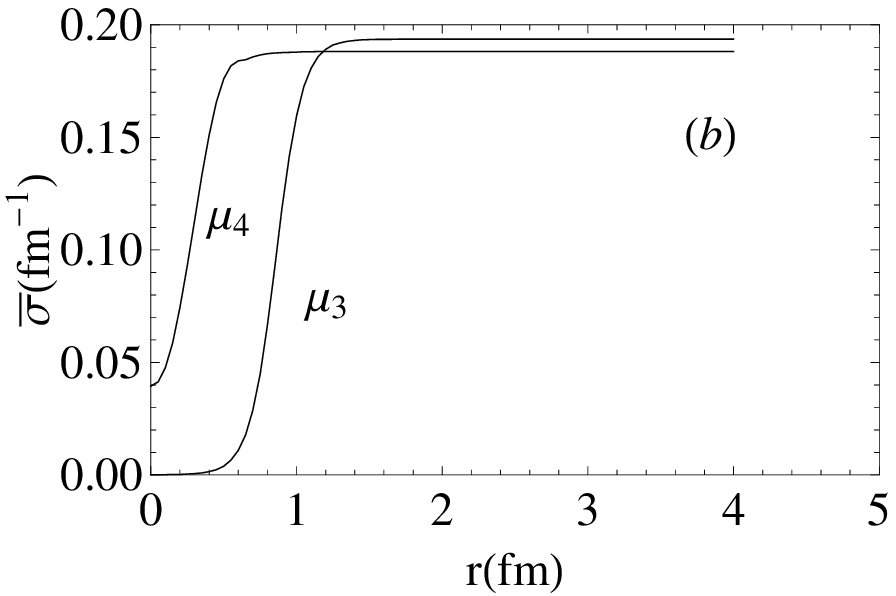}
\caption{The soliton solutions for the boundary condition II at
different chemical potentials. (a) $\mu_1=100MeV$, $\mu_2=200MeV$
and $\mu_{c1}=255MeV$. (b) $\mu_3=300MeV$ and
$\mu_4=350MeV$.}\label{f7}
\end{figure}

For different chemical potentials at fixed temperature $T=50MeV$,
according to the $V_{eff}$ as shown in Fig.\ref{f4}, the
corresponding soliton solutions could be solved and plotted in
Fig.\ref{f7}a and Fig.\ref{f7}b. It is clear that at
$\mu<\mu_{c1}$ it is a hadronic phase; at $\mu_{c1}<\mu<\mu_{c2}$,
it is in a sQGP phase; at $\mu>\mu_{c2}$ it is a quasi-free gas of
QGP.

From above discussions, it is clear that for the two different
boundary conditions, the physical meaning of the solitons are the
same: before deconfinement the solitons represent hadrons while
after deconfinment the solitons represent bound states of quarks.
For the two cases, the phase structure of deconfinement remains
unchanged and unique.

\section{The phase diagram}
\begin{figure}[tbh]
\begin{center}
\includegraphics[width=210pt,height=150pt]{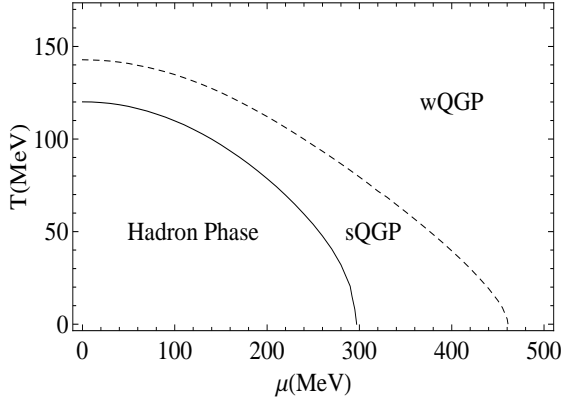}
\end{center}
\caption{The phase diagram of deconfinement in the FL soliton
model.}\label{f8}
\end{figure}

Now we are in a position to plot the phase diagram of
deconfinement phase transition in FL model. When the chemical
potential is fixed, one could obtain two critical temperatures.
Increasing the chemical potential to another fixed value, the
other two corresponding critical temperatures could be obtained,
and so on. The $\mu-T$ phase diagram could be plotted as shown in
Fig.\ref{f8}. The full line is the dividing line between hadronic
phase and deconfined quark phase. The dashed line further divides
the deconfined quark phase into the sQGP phase and the wQGP phase.

\section{summary}
In this paper the soliton solutions in FL model at finite
temperatures and densities have been thoroughly discussed for two
kinds of different boundary conditions. The physical
interpretations of the solitons in deconfinement phase transition
have been given. As in FL model the system is deconfined at the
time that the bag constant becomes zero, we indicate that the
solitons before deconfinement are hardrons, while the solitons
after deconfinement represent the strong coupled bound states of
quarks, which leads to the sQGP phase. The QCD phase structure
still needs lots of thorough investigations, especially for the
finite density areas, where the results are often model dependent.
Here we only present a possible phase diagram based on FL model.
The phase structure of deconfinement phase transition here is
qualitatively consistent with that of Shuryak~\cite{ref1}. The
whole phase diagram is divided into three phases: the hadronic
phase, the sQGP phase and the wQGP phase.

\begin{acknowledgments}
We are thankful to Hong Mao for helpful discussions on numerical
calculations. This work was supported in part by the National
Natural Science Foundation of China with No. 10905018 and No.
10675052.
\end{acknowledgments}


\begin{thebibliography}{}
\bibitem{ref1}
E.V.~Shuryak and I.~Zahed, Phys.Rev. C49 (2004) 021901.
\bibitem{ref2}
G.E.~Brown, C-H.~Lee, M.~Rho and E.V.~Shuryak, Nucl.Phys. A740
(2004) 171.
\bibitem{ref2a}
M.~Gyulassy and L.~McLerran, Nucl. Phys. A750, (2005) 30.
\bibitem{ref3}
D.~Teaney, J.~Lauret and E.V.~Shuryak, Phys. Rev. Lett. 86 (2001)
4783.
\bibitem{ref4}
P.F.~Kolb, P.~Huovinen, U.~Heinz and H.~Heiselberg, Phys. Lett.
B500 (2001) 232.
\bibitem{ref5}
D.~Teaney, Phys.Rev. C68 (2003) 034913.
\bibitem{ref6}
Z.~Fodor and S.D.~Katz, J. High Energy Phys. 050 (2004) 0404.
\bibitem{ref7}
F.~Karsch, PoSCPOD07 (2007) 026; PoSLAT2007 (2007) 015.
\bibitem{ref8}
M.~Alford, K.~Rajagopal and F.~Wilczek, Phys. Lett. B422 (1998)
247; K.~Rajagopal, Prog. Theor. Phys. Suppl. 131 (1998) 619;
R.~Rapp, T.~Schafer, E.V.~Shuryak and M.~Velkovsky, Phys. Rev.
Lett. 81 (1998) 53.
\bibitem{ref9}
D.T.~Son and M.A.~Stephanov, Phys. Rev. Lett. 86 (2001) 592; K.~
Splittorff, D.T.~Son, and M.A.~Stephanov, Phys. Rev. D64 (2001)
016003; J.B.~Kogut and D.~Toublan, Phys. Rev. D64 (2001) 034007.
\bibitem{ref10}
L.D.~McLerran and R.~Venugopalan, Phys. Rev. D49 (1994) 2233;
E.~Iancu and L.~McLerran, Phys. Lett. B510 (2001) 145; L.~McLerran
and R.~Pisarski, Nucl. Phys. A796 (2007) 83.
\bibitem{ref11}
T.D.~Lee, Nucl. Phys. A750 (2005) 1.
\bibitem{ref12}
K.~Yagi, T.~Hatsuda and Y.~Miake, Quark-Gluon plasma, Cambridge
Univ. Press (2005).
\bibitem{ref12a}
D.~Zwanziger, Phys. Rev. Lett. 90 (2003) 102001; Phys. Rev. D 70
(2004) 094034.
\bibitem{ref13}
R.~Friedberg and T.D.~Lee, Phys. Rev. D15, (1977) 1694; D16,
(1977) 1096; D18, (1978) 2623.
\bibitem{ref14}
R.~Goldflam and L.~Wilets, Phys. Rev. D25 (1982) 1951.
\bibitem{ref15}
M.C.~Birse, Prog. Part. Nucl. Phys. 25 (1990) 1.
\bibitem{ref16}
H.~Reinhardt, B.V.~Dang and H.~Schulz, Phys. Lett. B159 (1985)
161.
\bibitem{ref17}
M.~Li, M.C.~Birse and L.~Wilets, J.Phys. G13 (1987) 1.
\bibitem{ref18}
E.K.~Wang, J.R.~Li and L.S.~Liu, Phys. Rev. D41 (1990) 2288;
S.~Gao, E.K.~Wang and J.R.~Li, Phys. Rev. D46 (1992) 3211;
S.H.~Deng and J.R.~Li, Phys.Lett. B302 (1993) 279.
\bibitem{ref19}
H.~Mao, R.K.~Su and W.Q.~Zhao, Phys. Rev. C74 (2006) 055204;
H.~Mao, M.J.~Yao and W.Q.~Zhao, Phys. Rev. C77 (2008) 065205.
\bibitem{ref20}
U.~Ascher, J.~Christiansen and R.D.~Russell, ACM Trans. Math.
Software 7 (1981) 209.
\bibitem{ref21}
T.D.~Lee, Particle Physics and Introduction to Field Theory,
Harwood Academic, New York (1981).


\end{thebibliography}
\end{document}